\documentstyle[preprint,aps]{revtex}
\begin{document}

\preprint{\vbox{\baselineskip=15pt
\hbox{CGPG-93/12-1} \hbox{gr-qc/9312029}}}

\title{Extended loops: a new arena for nonperturbative quantum gravity}

\author{Cayetano Di Bartolo$^1$, Rodolfo Gambini$^2$, Jorge Griego$^2$
and Jorge Pullin$^3$}
\address{1. Departamento de F\'{\i}sica, Universidad Sim\'on Bol\'{\i}var,
Caracas, Venezuela\\
2. Instituto de F\'{\i}sica, Facultad de Ingenier\'{\i}a, J. Herrera y Reissig
565, Montevideo, Uruguay\\
3. Center for Gravitational Physics and Geometry, Pennsylvania State
University, University Park, PA 16802}

\date{\today}
\maketitle

\begin{abstract}

We propose a new representation for gauge theories and quantum
gravity. It can be viewed as a generalization of the loop
representation. We make use of a recently introduced extension of the
group of loops into a Lie Group. This extension allows the use of
functional methods to solve the constraint equations. It puts in a
precise framework the regularization problems of the loop
representation.  It has practical advantages in the search for quantum
states. We present new solutions to the Wheeler-DeWitt equation that
reinforce the conjecture that the Jones Polynomial is a state of
nonperturbative quantum gravity.

\end{abstract}

\pacs{04.65}

\narrowtext

The introduction of the loop represention has opened a new avenue for
the nonperturbative canonical quantization of general relativity.  In
particular it allows to immediately code the invariance under spatial
diffeomorphisms of wavefunctions in the requirement of knot invariance
\cite{RoSm}.  Also for the first time a large class of
solutions to the Wheeler-DeWitt equation has been found in terms of
nonintersecing knot invariants. It turns out however, that these
solutions correspond to degenerate metrics \cite{BrPu91}. If one wants
nondegenerate metrics one needs to consider knot invariants of
intersecting loops and solve  the Wheeler-DeWitt equation in loop space
\cite{Ga}. Solutions of this kind have actually been found
\cite{BrGaPuprl} and they are related with the existance of a
Chern-Simons state in terms of Ashtekar variables \cite{BrGaPunpb}.
One of these solutions turns out to be the first nontrivial
coefficient of a certain expansion of the Jones polynomial and this
led to the conjecture that the Jones polynomial may be a state of
quantum gravity \cite{BrGaPuessay}.

These results were plagued by regularization ambiguities.  Since
loops are one dimensional objects living in a three dimensional
manifold, they naturally lead to the appearance of distributional
quantities.  In particular the few knot invariants for which we have
analytic expressions require the introduction of regularizations
(framings) in the case of intersecting knots.  Some of them even require
them for smooth loops \cite{Wi,GuMaMi}.

This difficulty not only arises for the gravitational case.  In non
Abelian gauge theories \cite{MaMi,GaTr86}, and even in the simple case
of a free Maxwell field \cite{DiNoGaTr}, it is known that the quantum
states in the loop representation are ill defined and a regularization
is needed.

Loops are classes of closed curves that give the same holonomy for any
gauge connection.  They form a group under composition, called the
group of loops \cite{GaTr81}.  This group is not a Lie group, since
the composition of loops is only defined for an integer number of loops.
Recently, a completion of this group into a Lie group was introduced
\cite{DiGaGr}. The group of loops is included in this Lie group and can
be thought as a discrete subgroup. The elements of the Lie groups are called
``extended loops'' and are the mathematical basis of the new representation
we propose.

Let us discuss for a moment these ideas in the more familiar context
of Maxwell theory. The point of departure to construct a usual loop
representation is to consider the Wilson loop functional, $W_A(\gamma)
= \exp \oint_\gamma dy^a A_a(y)$, where $\gamma$ is a loop.

Since Wilson loops form an (over)complete basis of gauge invariant
functions (modulo subtleties we will discuss later) one can express
the wavefunctions $\Psi[A]$ in terms of Wilson loops and go to a
representation purely in terms of loops via the loop transform,
\begin{equation}
\Psi(\gamma) = \int DA W_A(\gamma) \Psi[A].\label{transform}
\end{equation}
The wavefunctions $\Psi(\gamma)$ are in the loop representation and in
this representation one can realize the gauge invariant operators of
physical interest of the theory, for instance the Hamiltonian, as was
discussed in reference \cite{DiNoGaTr}. It
needs to be regularized.  If one computes the vacuum, one gets,
\begin{equation}
\Psi_0(\gamma) = \exp(\oint dx^a \oint dy^b D_{ab}(x-y))\label{vacuum}
\end{equation}
where $D_{ab}(x-y)$ is the spatial restriction of the Feynman
propagator.
This quantity is also ill-defined due to the divergence of the
propagator where $x=y$.  So we have a representation where both
operators and wavefunctions have to be regularized (if one introduces
a regularized propagator, the expression (\ref{vacuum}) is the vacuum
of the regularized theory).

Consider now the quantities $W_A[X] = \exp \int d^3y X^a(y) A_a(y)$. If
$X^a(y)$ is a divergence-free vector density the $W_{A}[X]$'s are gauge
invariant. The $X$'s are the Abelian analogues of the extended loops
we will consider later on. If one uses $W_A[X]$ instead of the Wilson
loop functional in the transform (\ref{transform}) one ends with a
representation in which wavefunctions are functionals of transverse
vector densities. It is easy to check that this just corresponds to
the electric field representation of the canonically quantized Maxwell
field.  The Hamiltonian is well known. The vacuum of the theory is
simply given by,
\begin{equation}
\Psi_0(X) = \exp (\int d^3y \int d^3z X^a(y) X^b(z) D_{ab}(y-z).
\end{equation}
and is well defined without the need of a regularization.

We therefore see that by extending the idea of loop we have
several advantages: on the one hand we end up with a usual
representation in terms of fields (it is just the electric field
representation); on the other hand because we are using fields instead
of distributional objects (loops) the regularization difficulties
associated with the loop represntation are solved.

Another issue to be considered to complete the quantization is the
introduction of the inner product. There is no clear idea of how to
perform an integration in loop space and therefore there are no
natural candidates for inner products in the loop representation. By
going to the extended loops one can use usual functional integrals and
an inner product for Maxwell theory have been introduced this way
\cite{AsRo,DiNoGaTr}. With this inner product
the usual Fock structure and the interpretation of the excited states
in terms of photons can be recovered completely.

The intention of this letter is to outline the generalization of this
procedure to the nonabelian case, concentrating on the case of
gravity.

Let us start by proposing an extension of the notion of holonomy for a
non Abelian field similar to the one we introduced for Maxwell theory.
To this aim we rewrite the usual expression for the holonomy as,
\widetext
\begin{equation}
U_A(\gamma) = P \, \exp(\oint_\gamma A_a\, dy^a) =
1+\sum_{n=1}^{\infty} \int dx^3_1\cdots dx^3_n
       A_{a_1}(x_1)\cdots A_{a_n}(x_n)
X^{a_1\,x_1\,\ldots a_n\,x_n}(\gamma) \label{holonomy}
\end{equation}
where~$\gamma$ is a loop and the ``multitangents''
$X$ are defined by,
\begin{equation}
X^{a_1\,x_1\,\ldots a_n\,x_n}(\gamma)= \oint_\gamma dy_n^{a_n}
   \int_0^{y_n}dy_{n-1}^{a_{n-1}}\cdots \int_0^{y_2}dy_{1}^{a_1}
   \delta (x_n-y_n)\cdots \delta(x_1-y_1).
\end{equation}
\narrowtext
The advantage of rewriting the holonomy in this way is that we have
captured all the loop dependent information in the multitangents,
which behave as multitensor densities on the spatial manifold. The
holonomy therefore can be written in a very economical fashion as the
contraction $U_A(\gamma) = A_{\tilde{\mu}} X^{\tilde{\mu}}$ where the
indices $\tilde{\mu}$ are a shorthand for $a_1\,x_1\ldots\,a_n\,x_n$
and we assume a ``generalized Einstein convention'' in which we sum
from one to three for each repeated index $a_i$ and we integrate over
the three manifold for each repeated $x_i$. A repeated index with a
tilde also involves a summation from $n=0$ to infinity.

The key observation is to notice that if one substitutes in
(\ref{holonomy}) a multitensor density  $X^{a_1\,x_1\,\ldots
a_n\,x_n}$ (not necessarily associated with a loop) such that,
\widetext
\begin{equation}
\partial_{a_i} X^{a_1\,x_1\,\ldots a_i\,x_i\,\ldots a_n\, x_n} =
(\delta(x_i -x_{i-1}) - \delta(x_i -x_{i+1})) X^{a_1\,x_1\,\ldots
a_{i-1}\,x_{i-1}\,a_{i+1} \,x_{i+1}\,\ldots a_n\, x_n}
\label{dc}
\end{equation}
\narrowtext
the resulting ``extended holonomy'' $U_A[X]$ is gauge covariant under
gauge transformations. In other words, its trace is invariant under
gauge transformations connected with the identity. We will call it the
extended Wilson functional $W_A[X] = Tr(U_A[X])$. An important
difference is that the use of extended loops does not necessarily lead
to the construction of quantitites that are invariant under gauge
transformations not connected with the identity.  Therefore the
variables $X$ are able to capture more information than usual loop
variables. In particular, information of topological nature. This can
also be seen in 2+1 gravity, where usual loops fail to capture all the
information of the theory \cite{Ma}. As we mentioned before, loops
form a group. The quantities $X$ can also be endowed with a group
structure, and the presence of the extra elements (the ones not
associated with loops) allow to extend the group of loops into a Lie
group called the ``Extended group of loops''
\cite{DiGaGr}. The usual group of loops is a subgroup naturally
associated with the multitangents $X(\gamma)$. The group product
between multitensors is $(X_1 \times X_2)^{a_1\, x_1\, \ldots\, a_n\, x_n} =
\sum_{k=0}^n X_1^{a_1\, x_1\,\ldots\,a_k\,x_k}
X_2^{a_{k+1}\, x_{k+1}\,\ldots\,a_n\,x_n}$ and it naturally reproduces
loop composition among multitangents $X(\gamma_{1}\circ \gamma_{2}) =
X(\gamma_{1})\times X(\gamma_{2})$.

The extended Wilson functionals satisfy a series of identities
associated with the fact that the gauge theory is associated with a
particular gauge group (in the case of gravity $SU(2)$), which can be
explicitly written. In the case of ordinary loops these are the well
known Mandelstam identities

The extended loop representation can be constructed following the same
steps that led to the loop representation. Given a wavefunction in the
connection representation $\Psi[A]$ it can be transformed to the
extended loop representation via the ``extended loop transform'',
obtained by replacing $W_A(\gamma)$ by $W_A[X]$ in equation
(\ref{transform}).  One can also construct the extended
representation without invoking a transform directly by quantizing a
noncanonical algebra of quantities dependent on the extended
cooordinates, very much in the same fashion as in the usual loop
representation \cite{GaTr86,RoSm}, but we will not discuss it here for
reasons of space.

Some observations should be made about the resulting space of
wavefunctionals. First of all, they are functionals of the ``infinite
tower'' of multitensors of all orders. Moreover they are {\em
linear} functionals (since the extended Wilson loop is linear in the
multitensors). Wavefunctions must also satisfy Mandelstam identities.

Up to the moment the discussion has been generic, in the sense that it
could as well apply to any  gauge theory. We will now
particularize to the case of quantum gravity written in terms of
Ashtekar's new variables \cite{As},
\begin{eqnarray}
\hat{\cal C}_b \Psi[A] = \hat{E}^{a}_i \hat{F}_{ab}^i \Psi[A] = 0\\
\hat{\cal H} \Psi[A] =
\epsilon_{ijk} \hat{E}^{a}_i \hat{E}^{b}_j \hat{F}_{ab}^k \Psi[A] = 0
\end{eqnarray}
where $E^{a}_i$ is a triad, $A_a^j$ is the Sen connection and
$F_{ab}^i$ is the curvature constructed from the Sen connection. The
first set of equations is the diffeomorphism constraint which says
that the theory is invariant under diffeomorphisms of the three
manifold  and the last equation is the Hamiltonian constraint, which
corresponds to the Wheeler-DeWitt equation of the usual canonical
formulation.

In order to write these equations in the extended representation we
choose a polarization in which $\hat{A}$ is multiplicative and
$\hat{E}$ a functional derivative. We then consider the action of the
elementary operators on extended holonomies and rewrite them in terms
of the $X$'s, exactly as one proceeded with loops \cite{RoSm}. From
there one can obtain the action of any gauge invariant operator in the
extended representation. An important fact is that due to the
linearity of {\em all} wavefunctions in the extended representation,
any gauge invariant operator can only be a {\em first order}
differential operator. In particular the form of the constraints is
\cite{DiGaGr2},
\begin{eqnarray}
\hat{\cal C}_a(x) \Psi[X] &=& ({\cal R}_{ab}(x)
\times R^{b\,x})^{\tilde{\mu}}
{\delta \over \delta R^{\tilde{\mu}}} \Psi[X]  \\
\hat{\cal H} \Psi[X] &=& ({\cal R}_{ab}(x)
\times R^{a\,x\,{ \overline {b\,x}}})^{\tilde{\mu}}
{\delta \over \delta R^{\tilde{\mu}}} \Psi[X]
\end{eqnarray}
where $
R^{\tilde{\mu}} = {\textstyle {1 \over 2}}
(X^{\tilde{\mu}} + (-1)^n X^{a_n\,x_n\,\ldots \,a_1\, x_1})$,
$(R^{b\,x})^{\tilde{\mu}} = \sum_{k=0}^n
R^{a_{k+1}\,x_{k+1} \ldots
a_n\,x_n\,b\,x\,a_1\,x_1\,\ldots\,a_k\,x_k}$ and ${\cal R}_{ab}$ is an
element of the loop algebra such that the field tensor is given in
terms of the connection as $F_{ab} = {\cal R}_{ab}^{\tilde{\mu}}
A_{\tilde{\mu}}$. An explicit form for ${\cal R}_{ab}$ can be easily
written and the only nonvanishing components are of the first and
second rank. $R^{a\,x\,{ \overline {b\,x}}}\tilde{\mu}= \sum_{k=0}^n
(-1)^{n-k} R_c^{a\,x\,a_1\,x_1\,\ldots\,
a_k\,x_k\,b\,x\,a_n\,x_n\,\ldots\,a_{k+1}\,x_{k+1}}$ and the inversion
in the order of the last set of indices in this definition has a role
totally analogous to the ``reroutings'' at the intersections of loops
of the traditional Hamiltonian constraint in the loop
representation. The subindex c means take the cyclic combination in
upper indices. $\times$ is the product in the extended group of loops.

One can particularize the above expressions to the case when the
multitensors are the multitangents to a loop. The resulting
expressions {\em correspond to the usual constraints of quantum gravity in
the loop representation} \cite{Ga,BrPu92}.

What about solutions to the constraints? It should be pointed out that
since loops are a particular case of multitensors, {\em any} solution
found in terms of multitensors can be particularized to loops and
would yield in the limit a solution to the usual constraints of
quantum gravity in the loop representation (the limit could be
singular in some cases, for instance when loop expressions need to be
framed, as in the case of regular isotopic knot invariants
\cite{Wi,GuMaMi}). The converse is not necessarily true: given a
solution in the loop representation, it may not generalize to a
solution in the extended representation. An immediate example are the
unphysical solutions to the Hamiltonian based on smooth nonintersecting loops,
which find no analogue in the extended representation.

There exists a particular family of solutions in the loop
representation which {\em do} generalize to the extended
representation. In the connection representation based on Ashtekar
variables, the exponential of the Chern-Simons form built with the
Ashtekar connection is a solution of all the constraints of quantum
gravity with a cosmological constant \cite{Ko,BrGaPunpb}. When
transformed into the loop representation, the resulting wavefunction
is the Kauffman Bracket knot polynomial, which is a phase factor times
the Jones Polynomial. Through a close examination, it was conjectured
that the coefficients of the Jones Polynomial are solutions of the
Hamiltonian constraint without cosmological constant. Evaluating the
loop transform using perturbative techniques of Chern-Simons theory
\cite{GuMaMi}, explicit expressions for the Kauffman Bracket (KB)
coefficients can be found,
\begin{equation}
{\rm KB}_\Lambda(\gamma) = e^{\Lambda GL(\gamma)} (1 +a_2(\gamma)
\Lambda+a_3(\gamma) \Lambda^3+\ldots)
\end{equation}
where \hbox{$GL(\gamma)=g_{ax\,by}X^{ax}(\gamma)X^{by}(\gamma)$} is
the Gauss self linking number (\hbox{$g_{ax\,by} = \epsilon_{abc}
(x-y)^c/|x-y|^3$} is the free propagator of Chern-Simons
theory). $a_2(\gamma)$ and $a_3(\gamma)$ are coefficients of an
expansion of the Jones polynomial evaluated in $\exp(\Lambda)$ that
can be explicitly written as linear functions of the multitangents
with coefficients constructed from $g_{ax\,by}$. Through a laborious
computation it was shown that $a_2(\gamma)$ satisfied the Hamiltonian
constraint of quantum gravity
\cite{BrGaPuprl} and it was later conjectured \cite{BrGaPuessay}
that similar results may hold for $a_3(\gamma)$ and higher
coefficients.

These solutions can be easily generalized to the extended
representation simply by replacing the multitangents that appear in
the definition of the coefficients with arbitrary multitensors. It can
be checked that the resulting expressions are diffeomorphism invariant
and no framing problem arise. The remarkable fact is that due to the
simplification of the constraints that appears in the extended
representation, one can actually check in a straightforward manner
that $a_2$ is a solution as was already known. Moreover, the recently
obtained expression \cite{DiGr} for the third coefficient of the Jones
Polynomial, $a_3$, {\em can also be checked to be a solution of the
Wheeler-DeWitt equation} \cite{DiGr2} in the extended
representation. This adds more credibility to the conjecture that the
Jones Polynomial (in this case its extension to arbitrary
multitangents) could be a state of quantum gravity in the extended
loop representation. Notice that in all these solutions {\em no
assumption has been made on the domain of dependence of the
functions}. All solutions that were known in the loop representation
required restricting the domain of dependence of the wavefunctions
(say, to smooth loops or to loops with a certain number of
intersections).

Generically, the multitensors are distributional, as can be
immediately seen from the equation they satisfy (\ref{dc}). However
their distributional character is under control. As was shown in
reference \cite{DiGaGr} a generic multitensor satisfying (\ref{dc})
can be written as a linear combination of transverse multitensors
(which one can restrict to be smooth) times some well-defined
distributional coefficients. This has important consequences. In
particular the solutions considered above are such that the dependence
on the distributional coefficients drop off and they are only
functions of the smooth part of the multitensors. They are therefore
well defined. This was not the case in the loop representation, where
they needed to be framed to be well defined. This is because in order
to recover the usual loop representaion one needs to choose the
transverse part of the multitensors to be distributional.

The fact that the wavefunctions are well defined without
regularization ambiguities does not directly imply that the operators
in this representation are well defined. As in any functional
representation, the presence of functional derivatives may introduce
singularities that need to be regularized and renormalized. In the
extended representation, an obvious regularization problem is present
in the Hamiltonian constraint which involves a multitensor with a
repeated spatial dependence (as can be seen from (\ref{dc}) repeating
a spatial dependence involves a singularity). However, because the
singular nature of the multitensors is under control one can perform a
precise point-splitting regularization and it can be checked detail
if the coefficients of the extended Jones polynomial
presented above are annihilated by the regularized constraints or
not. This issue is currently being studied.

One can view the role of the multitensors in the extended
representation as configuration space variables of a canonical theory.
The conjugate momenta are represented by functional derivatives. This
suggests that there exists an underlying classical Hamiltonian theory
that under canonical quantization yields directly the extended loop
representation. This was unclear with loops, where the loop
representation could only be introduced through a noncanonical
quantization. For the Maxwell case this theory was studied
\cite{ArUgGaGrSe} and found to be equivalent to the usual Maxwell
theory. For nonabelian cases it is yet to be studied.

Summarizing, the extended representation presents practical
calculational advantages and offers new possibilities to precisely
regularize the theory and set it in a more rigorous framework without
losing some of the topological and geometric insights of the loop
representation.

This work was supported in part by grant NSF-PHY93-96246, by funds of
Conicyt and Pedeciba and by research funds of The Pennsylvania State
University.

\end{document}